# Algorithmic permutation of part of the Torah[1]

Grenville J. Croll

grenville@spreadsheetrisks.com

**A small part of the Torah is arranged into a two dimensional array. The characters are then permuted using a simple recursive deterministic algorithm. The various permutations are then passed through three stochastic filters and one deterministic filter to identify the permutations which most closely approximate readable Biblical Hebrew. Of the 15 Billion sequences available at the second level of recursion, 800 pass the a priori thresholds set for each filter. The resulting "Biblical Hebrew" text is available for inspection and the generation of further material continues.**

## 1. INTRODUCTION

The Torah decryption principles outlined in my earlier paper[1] suggest that T2, the 85 character section of text between the inverted nuns, might usefully be examined as a two dimensional array of 5 rows of 17 columns:

```
                              *                    Row No.

      ר מ א י ו נ ר א  ה  ע ס נ ב י ה י ו         0
      א ו צ פ י ו ה ר  ה  י ה מ ו ק ה ש מ         1
      פ מ כ י א נ ש מ  ו  ס נ י ו כ י ב י         2
      י ה ב ו ש ר מ א  י  ה ח נ ב ו כ י נ         3
      ל א ר ש י פ ל א  ת  ו ב ב ר ה ה ו ה         4
```

Given that there are slightly less than 85 factorial possible permutations of these characters it is infinitely improbable that any finite deterministic algorithm should permute the above lucid sequence into another lucid sequence otherwise than by design.

There exists the remote possibility that if the author or agency creating the original sequence is intelligent enough then it may be the case that algorithmic permutation of the original sequence might result in lucid derivative sequences coming to light either at all or at a rate that is higher than we would expect by chance.

In this paper I report the design and operation of a particular algorithm for permuting the above array of characters. I also describe some simple ways of filtering the resultant sequences such that the more "readable" are identified.

## 2. METHOD

Firstly, note that the five rows of seventeen characters can be interchanged in 5 factorial or 120 different ways. Secondly note that the characters in each row can be laid out left to right or right to left in 2 to the power of 5 or 32 ways. Thirdly note that we can skip through the 85 characters starting at the first character of the first row by using skip distances of 1, 2, 3, 4, 6, 7, 8, 9, ....42 etc to create a different derivative array of the same characters.

---

[1] Presented at the 2nd Conference of the Int. Torah Codes Society, Jerusalem, Israel, 6th June 2000

Skips that are multiples of 5 and 17 are not used because they are not co-prime to 85. The integers greater than 42 are also not used due to their redundancy given the combined effect of the left right symmetry of the row reversal step and the related symmetry of the skip step. (ie skip 84 gives the same result as left-right reversing all five rows and using a skip of 1).

Though my earlier paper gives a fuller background to the development of the algorithm, a few further points come to light upon implementation:

a) We work with rows rather than columns because 5! is tractable whereas 17! is not.

b) Note the significance of the characters comprising the central "pivot" (*) column.

c) To find the third Y-H-V-H in the equation 1839 - 3 = 1836 [2] (by switching rows 1 & 2 in T2) requires us to break the original sequence of characters comprising the Torah. In the event of ultimate design, this may be suggestive of consent to break the original sequence.

d) There are 32 unique skips. ie 32 = 42 - 8 - 2, and there are 32 row flips.

e) Sefer Yetzirah [2] verse 1, "With 32 mystical paths of wisdom..."

f) First character of the Torah is Bet, last is Lamed, which when reversed spell 32.

g) There are 32 actions of creation in Genesis 1 et seq.

Using some simple 'C' routines a sequence of characters has been generated for every combination of row interchange, read direction and skip distance making a total of 120*32*32 = 122,880 sequences, which are denoted the **Level One** sequences. The algorithm described above is referred to as the **third algorithm, zero offset.** The a priori decryption sequence in my earlier paper specifies an Offset operation which is not used in this analysis to reduce computation time.

Each of the Level One sequences has been used as the starting point for a recursive repeat of the above process. So we end up with 122,880 * 122,880 = approximately 15 billion sequences. These are referred to as the **Level Two, zero offset** sequences.

Clearly, given the volume of text produced by the above relatively simple algorithm, it is necessary to evaluate them using some form of automatic process.

## 3. SEQUENCE EVALUATION

To evaluate these sequences in terms of their readability, it turns out that Biblical Hebrew has numerous statistical properties which can be taken advantage of. Each permuted sequence was passed through a cascaded set of tests to determine in ever increasing likelihood whether or not the sequence belonged to the readable set. There are four tests. Three are acceptance tests, one is a rejection test.

## 3.1 The Quad, Pair, Triple (QPT) test

This test relies upon the obvious property of text in most languages that certain two, three and four characters sequences are more common than others. In English, we would expect a readable text to have its fair share of "THE", "BUT", "AND" etc, but not "ETG", "EFG" or "HVW" etc. Similar comments would apply to the frequencies of occurrence of two and four character sequences.

### 3.1.1 Generating the QPT parameters

---

[2] The Tetragrammaton appears 1839 times in the Torah. 1839 also happens to be the Neutron / Electron rest mass ratio, to 4 significant figures. The Proton / Electron rest mass ratio is 1836 to a similar approximation

Copyright(c) 2000-2010 Grenville J. Croll. All Rights Reserved.

Page 2

Using the Torah as the source document, three dictionaries were built comprising all the Quads, Pairs and Triples in the Torah. In addition, a count was maintained of how many times each Quad, Pair or Triple appeared in the Torah. Quads, Pairs and Triples that appeared five times or less in the Torah were ignored and deleted from the three dictionaries.

Three thousand separate 85 character sections of the Torah and three thousand separate 85 character sections of a Random Hebrew control were generated. For each 85 character section, the number of Quads, Pairs and Triples that appeared in both it and the Torah dictionary were calculated. These values were called Quadnum, Tripnum and Pairnum.

Clearly, some Quads, Pairs and Triples will be quite popular, whereas others will be obscure.

As a consequence of this, for each 85 character sequence a score was kept that was indicative of the absolute popularity of all the Quads, Paris and Triples appearing in within it. These values were called Quadscore, Pairscore and Tripscore. For example. if an 85 character sequence was full of obscure quadruples, then its quadscore would be low. Conversely, if an 85 character sequence contained the Divine name, then its Quadscore would be a minimum of 1839.

In order to evaluate the statistical differences between the values for Quadnum, Tripnum, Pairnum, Quadscore, Tripscore and Pairscore for the 3000 random and 3000 meaningful 85 character sequences, a simple multiple regression was performed. Quadnum, Tripnum, Pairnum, Quadscore, Tripscore, Pairscore were the independent variables. The dependent variable was either a 1 (Torah sequence) or a 0 (random sequence).

The R Squared for the multiple regression was 0.91 and the T-values (coefficient divided by standard error) for the variables were: 75, 9, 11, 15, 1.2 & 24. Given that there were just under 3000 degrees of freedom these results are highly statistically significant.

### 3.1.2 Running the QPT test

To evaluate any permuted sequence using the QPT test, the values of the Quadnum, Pairnum Tripnum, Quadscore, Tripscore and Pairscore were calculated for each sequence. These were then fed into the simple linear equation:

$$\begin{aligned}
\text{qptscore} = \;& \text{a\_constant} \\
& + a1 * \text{quadnum} \\
& + a2 * \text{quadscore} \\
& + a3 * \text{tripnum} \\
& + a4 * \text{tripscore} \\
& + a5 * \text{pairnum} \\
& + a6 * \text{paiscore}
\end{aligned}$$

where the constant and coefficients are obtained from the regression

Readable sequences have qptscores close to 1, random sequences have qptscores close to zero and the qptscore distributions for the two cases are given in Figure 1. For the QPT test and each of the following tests an a priori score threshold of 0.5 was arbitrarily chosen The probability of a random sequence having a qptscore > 0.5 is approximately 1 in 50,000.

When benchmarked against previously unseen sequences in Hebrew Bible text subsequent to the Torah, the QPT test perfectly discriminated between random and readable for Qptscore > 0.5. On a Pentium running at 200 Mhz, about 10,000 sequences a second can be generated and evaluated using the QPT test. The essence of the QPT test is its speed, as efficiently implemented it only requires three array lookups and the twelve flops comprising the regression equation.

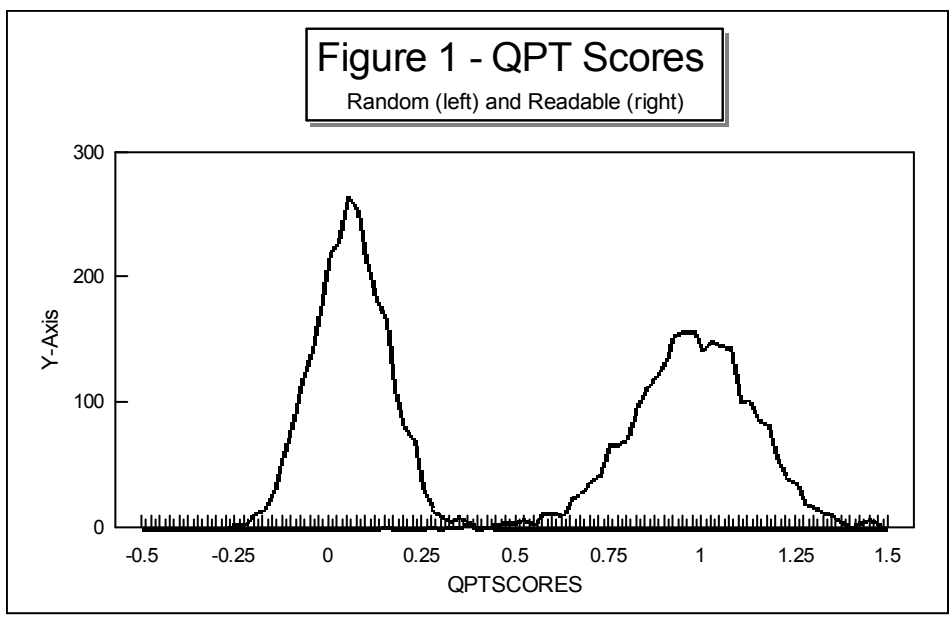

## 3.2 The Quads in Common (QIC) test

For many combinations of row interchange, read direction and skip, the permutation algorithm has a tendency to generate sequences which bear a strong resemblance to the original plain text of T2. A simple test is made to exclude those sequences which have more than five quadruples in common with the original plain text of T2.

The QIC test is a deterministic rejection test and is very fast.

## 3.3 The Word Test

Of course, readable Biblical Hebrew comprises words, not just Quads, Pairs and Triples.

For the word test, a Biblical Hebrew dictionary was required and such is available on the Bible Codes CD Rom from Computronic [3]. Some pre-processing was required to turn this dictionary into a list of words in alphabetical order, one per line. The Computronic dictionary is 39,640 words long and the maximum word length is 11 characters.

### 3.3.1 Generating the Word Test parameters

Again, 3000 Torah and 3000 Random 85 character test sections were extracted. For each 85 character section, starting at the first character, a series of dictionary lookups was performed to see how many valid Biblical Hebrew words started at each character position. The variable Wordnum was incremented each time a word was found. For example, starting at the B of BRA$YT, we would count three words BR, BRA and BRA$YT. Having found three words starting at the first character B, the character pointer was moved to the second character R and the number of words starting with that particular R was also counted. All 85 characters were examined in this manner, wrapping round back to the beginning of each sequence, as if the characters were on a ring.

For a readable sequence, it is generally the case that every character in the sequences is covered (or "spanned" by a character within a valid word. So, for each of the 85 characters on the ring a count was maintained of how many times each character was covered or "spanned" by a word. From the data maintained about each sequence, the independent variables Maxspan (character position with the most valid words passing over it), Minspan (similar), Unspan (total number of characters not covered at least once by a word) and Totspan (total number of times characters were covered by words) were created.

A multiple regression similar to the QPT test was again performed using 1 (Torah Sequence) and 0 (Random Sequence) for the dependent variables. The resulting R squared was 0.90 and the t-values of the variables were 7, 29, 32, 19 & 15, indicating a high degree of statistical significance for the model and all its independent variables.

### 3.3.2 Running the Word Test

To evaluate any permuted sequence using the Word Test, the five independent variables wordnum, maxspan, minspan, totspan and unspan are calculated from the permuted sequence under examination and Wordscore is calculated:

$$\begin{aligned} \text{Wordscore} = \quad & b\_constant \\ & + b1 * maxspan \\ & + b2 * minspan \\ & + b3 * totspan \\ & + b4 * unspan \\ & + b5 * wordnum \end{aligned}$$

On a 200 Mhz Pentium, the word test takes about one second, however it only needs to run once every five seconds.

Examination of the features of Biblical Hebrew unveiled during the development of this test showed that there are multiple word possibilities at every character position. In English there are far fewer word possibilities within a readable sequence as cursory examination of this sentence will reveal.

## 3.4 The Path Test

A readable Biblical Hebrew sequence contains a sequence of words starting at the first character and ending at the last character. Ie there is some combination of words that enables the sequence to be scanned from one end to the other in an unbroken list of words.

Given the variety of valid words that appears in any readable Biblical Hebrew sequence there are many combinations of ways that words could be abutted or concatenated such that the sequence scans from one end to the other. The ease with which a sequence of abutting words can be found might give some information regarding the potential readability of the sequence.

Finding a valid word sequence is a simple variation of the Knapsack problem. The Knapsack problem is NP complete and so to calculate the ease of concatenating words together a brute force approach was chosen.

There are 85 possible start points for a readable sequence of characters on an 85 character ring. A thousand start points were chosen at random, ie about 12 attempts were made to string a sequence of characters together from each of the 85 start points. For each of the thousand start points, approximately 1500 random attempts were made to increase each sequence by adding on words that directly abutted. A list of valid words and their offsets was calculated for each permuted sequence and it was from this that longer strings sequences were created.

Whilst this stochastic process was taking place, a count was kept of the number of strings that were of length 25, 45, 65 and 85 chars. The number of iterations to achieve the first 85 character sequence was noted, and the maximum sequence length also noted.

The above was performed for Torah and Random sequences and a regression performed in the usual fashion. The R-squared was 0.89 and the t-values for the variables were 67, 7, 8, 5, 5 & 9. Thus the Path test involves calculation of the following values:

$$\begin{aligned}\text{Pathscore} = \ & c\_constant \\ & + c_1 * maxpara \\ & + c_2 * num25 \\ & + c_3 * num45 \\ & + c_4 * num65 \\ & + c_5 * num85 \\ & + c_6 * iterations\_to\_85\end{aligned}$$

The Path test is therefore a measure of the degree of difficulty in stringing together words to make longer sequences, bearing in mind that there are on average a hundred or more word choices in any readable 85 character Biblical Hebrew sequence.

The Path test takes approximately 10 seconds to execute, but only runs very occasionally.

## 4. PRELIMINARY RESULTS

Approximately 15 Billion Level Two sequences have been generated. This took three 200 Mhz Pentiums working in parallel approximately 10 days.

For each sequence the QPTscore was calculated. If it was above 0.5 and the number of Quads In Common was less than 6 then the Wordscore was calculated. If the Wordscore was greater than 0.5 then the Pathscore was calculated. If the Pathscore was greater than 0.5 then the permuted sequence was printed out together with some diagnostic information.

Approximately 850 sequences passed all four tests and are available for inspection. A small random sample is included in Appendix A. Commercial translations of earlier sequences are given in Appendix B.

## 5. ENGLISH & HEBREW CONTROLS

The Park Miller minimal standard algorithm was used to generate 85 character sequences with the correct Biblical English letter distribution. These were evaluated by the English equivalent of the QPT test. Although the random English sequence generator ran for a period of many days, no English sequence achieved a Qptscore of greater than 0.5.

The minimal standard algorithm was also used to generate 85 character sequences with the correct Biblical Hebrew letter distribution. These sequences passed the four tests described above at a similar rate as permutations of T2

It remains to be seen if permutations of T2 are more interesting than random 85 character sections of Biblical Hebrew generated using a pseudo random number generator.

## 6. CONCLUSION

Biblical Hebrew has the unusual and previously unknown property that it is relatively easy to algorithmically generate long strings of characters that pass some quite severe tests for readability. This is no doubt due to the consonontal nature of the language and, perhaps, the oft noted[5] mathematical precision of its grammar.

A small part of the Torah was permuted using a simple algorithm into strings that bear some of the characteristics of readable Biblical Hebrew. Although it is clear that the sequences generated, some of which are reproduced in Appendix A, contain sequences of words, it is extremely unlikely, but not impossible, that one of the sequences might be lucid. Commercially produced translations of some early sequences are given in Appendix B.

A wider search for lucid sequences could continue by using an increased depth of recursion, by including the offset or rotation step, or by using variations of algorithm three. The score thresholds could be increased from 0.5 to reduce the amount of potentially readable material produced.

Searching for lucid sequences could be done on massively parallel architectures, on the Internet or on easily designed, but expensive, custom VLSI. The process of reviewing the material produced would be tedious, however it may be possible to further reduce the number of sequences for human review by developing a Biblical Hebrew Grammar checker or further more advanced technologies.

Speculatively, T2 might be some sort of index for the whole Torah.

Suppose that at some point in the future, using some variation of Algorithm Three or some other algorithm, that T2 is permuted into at least one interesting and lucid sequence. If we have recorded the particular sequence of actions - row permutations, left right flips and skip sequences etc for each recursion level and each lucid sequence, then these parameters might be a key sequence which could then be applied to the whole Torah.

Note that the Koren edition of the Torah, less the inverted nuns is 304,805 characters long. These characters can be laid out into a 5 * 60961 array. It would be quite natural to be tempted to apply any interesting T2 key sequences to this array. And this is entirely possible because the number of rows is the same as in T2.

## 7. ACKNOWLEDGEMENTS

I would like to thank Dr Moshe Katz for his efforts in organising the ITCS 2000 congress and the other delegates including Professor Robert Haralick, Professor Eliyahu Rips, Professor Daniel Michaelson, Mr Harold Gans, Mr Art Levitt, Mr Nachum Bombach, Rabbi Y. Zilber and Rabbi Y. Spielberg for their hospitality. I thank Professor John H. Conway for his comments on ineluctability at my presentation of this work at ANPA 2005, Wesley College, Cambridge, UK. This paper is dedicated to the memory of my father, William John Samuel Croll, (1927-2010), "…..a superb professional Engineer".

# APPENDIX A

Sample permuted sequences produced from t802.cpp on 15<sup>th</sup> January 2000, which implemented recursion level two of Algorithm Three, Zero Offset. The seven digit number to the left of each sequence is a numeric identifier obtained from the sums of the squares of the ascii values of adjacent character pairs. No attempts have been made at translation of these sequences.

```
3979074  להשייהורהונהערסברמכמהארפורהיאייאחיויומיפניכוהילביוכרתשוהנאוותסאבינואבנומשנפשביהיהימא
4212252  אהוואמויצרשויהיינותריאבננבכיקרואבוסתפולהיהיכמלביהבאסתפניהנכהיהואשמוחישיהראנשיוממימור
4219887  מפשהוויחייהיהואהרנצבנאוויאאכמריהוורומנסוהנאהלימיהוספהשניכיפישהקתנבכירברומלאבאמיהויערשויי
4332664  הבוראילכנאוראניליהמויינתספפכמהיכירועציומבנוואבוורקוסבביתהבמאשיהשמנהורסבראהחיהההנהאיהמי
4332868  צנפרינאקבאוויכתיהכסהואיליוובאשריוספיומהובאיוישההרהממהחרביינהאלועראפמהכיומשנבוומשנגא
4401851  פרנהשיחהויבנשוכובסיספלהעונוימאמההומכרלהמבאנהובניהואישאקאכתרביויאשרבמפרוויניאצהיהואי
4403364  יהלוישכניהאיוהבשלחכנרבבפרצוישסהספמפהיבאהיוויההינאיאבכרנוותאיפעורנואומשרמישומאמממה
4495420  מקשלכומנאספוליההאחיראנוביהיהיובהמנעציאמרורמהייומנוכיבאניהיאאובוהותפתבנהוהוכפישוסוירב
4501056  ינובאביאהבההאשהיההילעימנושואתמטספהצלההאיררירפריחיוונאבינורמכהנבוסבואכיימושנופורשמ
4590143  רמאויצפנבבסעננברנוקראלאונגיאומהמביהשמהוהישרתויהיפויכינובהוהריוהואלובישכמפיהכיהנסהאחנאי
4590747  ויהיאבותהומראביפמכרבמהייארנוכיאלהוהקונהזההאניושראומעהונבנסעוספייפצהיניכיבנשמאלהווובמר
4666699  לאירבניווויישכננכסנהוונספייהיהבמליאשבאוכריאבנונהממקמהכייהשמנופריספיוהיאתרמצאשויוהאחוערב
4666831  וויש יאבירכופריהבינימסטרוחתומרמופשעההאלבויההיהואמנהומכביאלנויש ומהנהניהכיהיאבאצתשיאברפנ
4719926  כיהיהבוריהעלהשביכיאמיואשתהויהיאוהואלונצמפושוורוהנווירשרקבימאנספרהנמהנספרבניחופהאאמכ
4720356  ראוכובנשמאלהמשואמינויהאאמנוכיאלפיהיינתהינריהבימוכרברומהיאריושמהוצפנבנסעוספיוקבה
4775957  ובנראהוהיהואלמאקמפלנסעברהיהרומארכויובושמשניפצוהיתהאניכנפמחהיתוהבאיכהירריהיויננוסאיומרביהיישה
4779114  שפוופהההוננמקאבאשתהופראתיהיבחריהיוליוכהיויועוהירייומארהמרהכנאאממסטסבננויויבוכויבהאוהלשיבירמ
4840707  יכובבאהלשיברומשפוופהההוננמקאבאשתהופראתיהיבחריהיוליוכהיויועוהירייומארהמרהכנאאממסטסבננויי
4841087  ובנההנסעשובארהויונצפילפיביהאלמיכמיבשמואמרוכיהכמקומתאוהויאשרבברמאנראהסנייהחיונשה
4904230  שבושיהואניהכהוכספההנואביאשיהיונהוארומהנצנשונמתוריהמפפחסירעמאהארהלכיבינולויאיקרעיבוי
4904235  אתמכריהשמנהחיויכנהואביבוסטאביפויכיאישנוהספאעצייהומיהממקרהמירהמרנבפסלויאיבוהאמהאלבהובינו
4952268  יוורישמולאינהיויהונסכופייומאמראמהבהוההנהוקבנסיהארניבצורמפיהאנצייכתובוכרבימעשהאבנהדיאלאיפוש
4953824  מלאותאוהויהיהאפהוקמושמויפלוכיננבומכוביאיואמריהבינויהסעהיהוישומרכיבצפיארנישנאיוש
4998978  ממהרפואונולשלבתסתיאוסטפושאיסייונומאהושאתרתאמהיבישאיוונהומעצינייברקאבכהפרכהיוכלבינואיוומיבנהה
5000039  ונצפיהחנביהאישהמוירבההויהיההואאפאמקומשונסעומנטשייכובתובאלפלביהלהביהואמאיינסברהינורא
5027503  מפמנסטואמראיכההיהיבהרבעברמואהוישאריבערבכיהויומתאוההישאמהסניהחוונשהובנבהיהוחיונשנבנהסעהיהואריהונצפילפיביהאליי
5028306  פהאבניוההנהסלהשספקפפמתיבויהונשכארההמבניהויהוהאמילויכיהנצארמניכואישורוורישאבביהואספימ
5063266  פמשמונההאריישיויניסטיבוכייהבריהקוטבואיאמינלבכמאהההונהנסעברהייפיראיאניכוריהמאלפרשומא
5063390  רופויישאאהכוהנהנהמאויתמיפצבינכנייוינההוריופימרמוכיהסבימואהיהיבארחלשמובעשהקיוואראהוס
5113381  נבשהבביליישמוישירההויהיאאתהעמרמסטופהיוויורכיההואניהואיושנואהואויהסיפמהוהכברבינונהכמלבנמחיהימאפ
5113419  וצירחישופיבאמאישנשמתההאקמנאלבנבניהיוהיינההיכלהיארםמווארמווורפריהומהויהויהבכסמסכפרנעאביביהו
5153190  מאספאתהונאההשמימוקיפושפופיאמרלבויוישרבויצוננאנמבאשיסנהיובהוובבזהויכיכיהרבביעבירעכיאילמה
5153318  נירואיהיוישמויעהיתבאויוכמיאמהתהמרירהאובתפשלפיבנכנואשההוריהקמאנהנההרולביאבוסשיפכצ
5192364  ימאלנכופהיתהונכילהשוהרימוניסיהואהיינושרהובריובורופימקרבעקימוייבראהויוהאאאסשצגנבאשוואנהוהשהיה
5192722  וסברכיראנכיפאבבואאהויחיונסיההמנהיהניומנצולומנאשיתההימיהיכונמימייהייאשייהיהעמעליסטואבבברריפריקרבניפתוהו
5231426  פרהנוהוהבאאלכסטומיסראומשפיוונברתומישכיראצהותיהאייבמיהיוויהבויבונוהנוקיובבמלימוהאאפואניהה
5233370  ובנהרבינשוטוונרארליהצפיתהלאפיבאלפהרכימואירהרכירבוהואבנהאראמקימיכוכרכוהותהייישמפכפאוסמונסעהיחבנינוש
5275910  לבינאהנכייההההחזחייהנצאמוצענונאהמכאפההוהורולילאיש ומראשנטסיבכבטוהיימויריהינוקיפתרפשיכצפמרהואימשטבורואבש
5276022  פסוכלקבהיווישאריסבוהשאריהאיההאמנוהוהוההכהעורפימבאנויהייאאמומזעאתיההירחימנפויבאמרוהוכעינייוכשהיובשיני
5332683  וכהלמשוטובעויהיהמירווישאאמיהרמאנההרהגהורפנויניוהביהנאסטוהורמנוכפסימבאומסרוהרמנסבימוגעואנופויקלויארא
5333766  ינהמאיפסאריוסכפהאאווישאאוישנבבריוסיפיהאראבאהוסרההואהחיהעיומיסהנכממנלירהוסיההבמהקצחנתיהכומחוברמנעל
```

```
נשרפוכנעמיקהשכחנההאינואבסימאמיהוהמיההיוהיוונפלמוארהבויומברתיהיניהשצבסטופטראיורבלכואהבוא   5402227
אימרויומופצפאחיכפחנרסעוארבוהויאהיאהסקכותבנולראשויבויראהאלפמהוהנבהנההנגשינאינייכבורמסי    5402266
איבשהיהשבונוזוסטפוהלאנאנכויקויבאתליפומיהמראסרנבההערוהמכיימוהיאנשפרהואירצמומבשבכחיונה   5461965
יונשמונסעוכיהירשהובנבוששאמריהואהמאניבראהוהויאשירויומכמקומתאסניפצבבררוהיאנחזהילפיהיאלפיכ   5463894
לפסכירנהבנומנוהארחירהסמראיההמבויצילוסימעועיאשירובתיאיפומוהוהאומנוכאיהסקשנפהאהבנייהויכ   5524467
וראיתיצירהושמבנאחהבויומיורומהפקומרהניפהבאמפבאיהוברינרנוטבעיההאשווישנסיהוכאלהיכלמאם     5526366
נחפרותהצוויהיויהרסהחרקירהבאנומוירומלבמלאויומעליאבהראשכאשכפהבמכובניניוהנויששיספהבמושאבניא   5615669
סבמואופישקליוראוכהלשמובעביהיוממרוהמירווייאאחרהאנהנהררופניוייההבהשיצינכתיבאומטוהרמונכפימ   5615979
יכנהרהמולהביהיהיהקינבנויישהוטבווהמימבויתאכפפנומנאריאומבטוסצאהייוושנשמובערהלכהארבהרסרפואישומח   5750022
פיופלאאמריהיחירישארובנעטהבוחותההוהיהנהיהיאהנקוינטהוברפוצאריארונכיטמלכמוהשמויהמברהנביכי   5750862
וארירכיהיהיכושוביהיהמיכותביהיסמוסלמוסיפיבאמלפנהיבעיפראתוינצואשנכתומכחיהאמרוהובניומישאנסעוהרנבראהה   5959370
יהלקחסבימיירכהרנהפיניראמואובמוהאאורהאראלהויכסומצבישתימיהיכוננבנאנהרושאנעופשופי    6002577
```

# APPENDIX B

Commercial translations, from Biblical Hebrew into English, of permuted sequences produced by program t702.cpp on 3rd September 1999. T702.cpp implemented Level Two of Algorithm Three, Zero Offset, but did not contain the Path Test.

214687792

1.
נא פכ ה-י ההה הכוי ע שרפיו ונסנ מש ומריו מקהו אבאו ורי
ירשב ונל ה-י מרא יאו מלצ חס המיאב נבוכיפתי שא היה ינבו יב

? and father were awakened and lifted up/exalted we fled there and withdrew defeated/smitten ? [by] G-d's hand ? in me and my son there was fire ? says the Aramean image [of] G-d in terror and woe upon us my flesh

2.
נא פכ היה הה הכוי ע שרפיו ונסנ מש ומריו מקהו אבאו ורי
ירשב ונל יאו הימרא מלצ המי חס אב נבוכיפתי שאהי הינבו יב

Jeru and father were awakened and exalted there we fled and retired ? cheated ? there was a hand ? in me and her sons Yeho'ash ? it went towards the sea says [the] army photographer and woe upon us my flesh

214702352

ו אבאו ורי ירשבנו נליאו ה-י מרא מלצ המיאב חס נב וכי פתיש
א היה ינבו יבנא פכ ה-י ההה הכוי ע שרפיו ונסנ משו מריו מקה

a collaboration will strike us in terror says the Aramean image [of] G-d/army photographer and the tree and my flesh will fire and father ? will be awakened and exalted/lifted up and there we fled and retired/withdrew ? smitten/defeated ? [by] G-d's hand ? and my son was ?

214724112

יו ונסנ משו מרי ומקהו אבא וורי ירשבנו נליאו ה ימרא מל
צ חס המיא בנבוכיפתי שא היהי נבוי בנא פכ הה הכוי ע שרפ

why did [the] Aramean ? and the tree and my flesh saturate father and awaken him and exult him and we fled there ? withdrew ? smitten ? there was a hand ? it will be understood fire ? fear/horror/terror said ?

214778000

הימרא מלצ חס המיאב נבוכיפתי שא היהי נבוי בנא פך היה
הה הכוי ע שרפיו ונסנ מש ומריו מקהו אבאו ורי ירשבו נ ליאו

there was a hand ? [which] will be seen/understood that there will be fire ? in terror says [the] army photographer and the ram/lord ? and my flesh fired/? and father were awakened/raised and tricked/deceived we fled there and withdrew ? defeated ?

31908976

ו הצמ פסכ מה יבה ולבי נ רסיימ ו רשאו שאו אי שרפ השמ נתנה
וח הכי יאקביאוא יהי רהנ מויומוי לכ וב הב נניא

He who allowed Moses withdrew/retire ? fire and who ? tortures ? will spend/outlive ? they found silver/money ? there is nothing in her all/everything in him and he beheaded daily load/burden river shall be ? defeated ?

92536064

ניבי כלה ררא המא מרי סחפ פמ ה-י ימו ותמ נושנצבנ המור א
והנ ויהי שיא ביאוכ הפסכו הכי הנא והיש ובשי וב י ערקי אלו

? were cheated ?changed died and who G-d ? straightened and shall exult/lift mother ? he they will come to understand and not tear ? where they dwelled ? the boat shall be smitten/defeated and her silver/money as an enemy [of] one man ?

92576672

אוה יש ובשי וב י ערקי אלו ניבי כלה ררא המא מרי סחפ פמ ה-י
מו תמנ ושנצבנ המו רואה נ ויהי שיא ביא וב הפסכו הכי ה-י ־נ

G-d ? straightened [and] shall exult/lift mother ? they will come to understand [that] he will not tear ? the place where they dwelled he is a gift ? G-d will smite and her silver/money wherein there is animosity/feud/malice will be ? the light and which ? we will die ?

92589536

בן המו ראוה נ ויהיש י אביא וב הפסכו הכי הינאו ה-י שובש
יובי ערקי אלו ניבי כלה ררא המא מרי סחפ פמה ימו ו תמנו שנצ

upset./distort G-d and he will smite the boat and her silver/money wherein there is animosity/feud/malice ? there will be ? the light and whatever ? and we shall die ? and whoever ? is straightened will exult/lift mother ? he will come to understand and will not tear ?

92605056

ביא וב הפסכו הכי הינאו הישו וב י ערקי אל וניבי כל
הררא המא מרי סחפ מהימו ו תמנו שנצבנ המור אוה נ ויהי שיא

go [and] you shall understand not to tear ? where he dwelled ? and the boat shall be smitten and her silver/money in him ? there will (will = plural) a man ? he was cheated ? and we shall die ? and from them ? straightened exult/lift mother cursed her

92647232

סחפ פ מהימו ות מגו שנצבנ המור אוה נ ויהי שיא ב יאוב הפ
סכו הכי הינאו הישו וב י ערקי אלו ניבי יכל הררא המא מרי

here come here ? a man will be (will be = plural) ? he was cheated ? as well as a mark/character from them ? straightened shall be exulted/lifted mother cursed her go between him and do not tear ? where he dwelled ? and the boat will be smitten and [the] throne